\documentclass[12pt]{article}
\usepackage{amsmath,amsfonts,amssymb,bbm}
\usepackage{tensor}
\usepackage[T1]{fontenc}
\usepackage[utf8]{inputenc}
\usepackage{lmodern}
\usepackage{authblk}
\usepackage[disable]{todonotes}
\usepackage{color}
\usepackage{multicol}
\usepackage{cite}
\usepackage{collref}
\usepackage[margin=3cm]{geometry}
\usepackage{hyperref}

\newcommand\td{\text{d}}
\newcommand\cO{{\cal O}}
\newcommand\cA{\mathcal{A}}
\newcommand{\cC}{{\cal C}}
\newcommand{\skyp}{{\cal I}^+}
\newcommand{\skym}{{\cal I}^-}
\newcommand{\bz}{\bar{z}}
\newcommand{\pu}{{}}
\newcommand{\gr}{{}}

\newcommand{\p}{\partial}

\newcommand{\be}{\begin{equation}}
\newcommand{\ee}{\end{equation}}

\def\nn{\nonumber}
\def \ga {\gamma_{z\bz}}
\def \gai {\gamma_{z\bz}^{-1}}
\def \sg {\sqrt{-g}}
\def \bw {\bar{w}}
\def \gawi {\gamma_{w\bw}^{-1}}

\allowdisplaybreaks[1]
\newcommand\email[1]{\thanks{\href{mailto:#1}{\nolinkurl{#1}}}}

\author[a]{Pujian Mao\email{maopj@ihep.ac.cn}\,}
\author[a,b]{Hao Ouyang\email{ouyangh@ihep.ac.cn}\,}
\author[c,d,e]{Jun-Bao Wu\email{junbao.wu@tju.edu.cn}\,}
\author[f]{Xiaoning Wu\email{wuxn@amss.ac.cn}\,}

\affil[a]{\,Institute of High Energy Physics and Theoretical Physics Center for Science Facilities,
Chinese Academy of Sciences, 19B Yuquan Road, Beijing 100049, P.~R.~China}
\affil[b]{\,School of Physical Sciences, University of Chinese Academy of Sciences, 19A Yuquan Road Beijing 100049, P.~R.~China}
\affil[c]{\,School of Science, University of Tianjin, 92 Weijin Road, Tianjin 300072, P.~R.~China}
\affil[d]{\,School of Physics, Beihang University, 37 Xueyuan Road, Beijing 100191, P.~R.~China}
\affil[e]{\,Center for High Energy Physics, Peking University, 5 Yiheyuan Road, Beijing 100871, P.~R.~China}
\affil[f]{\,Institute of Mathematics, Academy of Mathematics and System Science, Chinese Academy of Sciences, Beijing 100190, P.~R.~China}

\title{\bf New electromagnetic memories and soft photon theorems\\}
\date{}

\begin{document}

\maketitle
\thispagestyle{empty}

\begin{abstract}
In this paper, we present a new type of electromagnetic memory. It is a `magnetic' type, or B mode, radiation memory effect. Rather than a residual velocity, we find a position displacement of a charged particle induced by the B mode radiation with memory. \pu{We find two types of electromagnetic displacement (ordinary and null).} We also show that the \pu{null} electromagnetic memory formulas are nothing but a Fourier transformation of soft photon theorems.
\end{abstract}


\section{Introduction}
Gravitational waves are observed by geodesic deviation of nearby freely falling observers. An \pu{interesting} of gravitational waves called `bursts with memory' will induce permanent relative displacement of nearby observers. Such effect is the well known gravitational memory effect. It was first noticed by Zel'dovich and Polnarev \cite{memory} in the study of scattering of stars in linearized gravity theory. The possible detection of gravitational memory effect was proposed in \cite{Braginsky:1986ia,1987Natur}. Soon afterwards, Christodoulou reported a non-linear memory effect based on a rigorous analysis of the asymptotic behavior of gravitational fields \cite{Christodoulou:1991cr}. However it got comments shortly that the non-linear memory is basically the gravitational fields produced by the burst's gravitons \cite{Wiseman:1991ss,Thorne:1992sdb} \pu{(see also \cite{Tolish:2014bka} for further clarification)}.

Apart from the interest in gravitational wave experiments, the gravitational memory effect obtains
renewed attention in purely theoretical point of view recently. Strominger and Zhiboedov \cite{Strominger:2014pwa} discovered a deep connection between gravitational memory effect and Weinberg's soft graviton theorem \cite{Weinberg:1965nx}. The gravitational memory formula is nothing but the Fourier transformation of Weinberg's formula for soft graviton production. Moreover, accompanied with an earlier discovery \cite{He:2014laa}, a triangular equivalence has been found. The precise ingredients of the three corners are BMS super-translation \cite{Bondi:1962px,Sachs:1962wk}, Weinberg's soft graviton theorem and gravitational memory effect.

Recent development in scattering amplitudes shows that the universal soft graviton formula goes beyond Weinberg's pole formula. It contains also next-to-leading orders in the low-energy expansion (see \cite{Cachazo:2014fwa} for a comprehensive discussion). If one believes the triangular relation to be valid in general, the new universal soft graviton formula should hint new gravitational memory which was proposed in \cite{Pasterski:2015tva}. The new gravitational memory is a spin memory effect. It will induce a relative time delay between
orbiting light rays. The spin memory effect was confirmed in post-Newtonian approximation very recently \cite{Nichols:2017rqr}.

As a theory with gauge invariance, electromagnetism often provides some simpler analogues of gravitational effects. This also happens in memory effect. Bieri and Garfinkle found an electromagnetic analogue of gravitational memory effect \cite{Bieri:2013hqa}. \pu{In analogy with the gravitational wave memory where two memory effects exist, namely linear and non-linear (or, ordinary and null) effects, they also found two types of electromagnetic memory (ordinary and null).} The electromagnetic radiation with memory will cause a change of the velocity (a `kick') of a charged particle. \pu{An experimental proposal to detect the electromagnetic memory effect was suggested by Susskind in \cite{Susskind:2015hpa} recently.} On the other hand, the triangular relation in gravitational theory starts to emerge in electromagnetic theory with certain adaption. The leading soft photon theorem was identified as the Ward identity of large $U(1)$ gauge transformation \cite{He:2014cra}. The equivalence between `kick' memory formula and the formula of leading soft photon theorem was also noticed in \cite{Pasterski:2015zua}. It is known long time ago that the universal properties of soft photon theorem go through the
next-to-leading order in the low-energy expansion \cite{Low:1954kd,GellMann:1954kc,Low:1958sn}. Correspondingly, a new electromagnetic memory seems to exist according to their counterpart in gravitational theory\footnote{An analysis on the possible memory effect associated to sub-leading soft photon theorem was also suggested by Strominger in a very recent lecture note \cite{lecturenote}.}. This is precisely what we will confirm in this paper.

We find a new type of electromagnetic memory. \pu{The new electromagnetic memory also includes the ordinary and null effects.} It is a magnetic type of radiation memory and will cause a position displacement of a charged particle. Interestingly, the new memory effect \pu{has the same formula as} the Aharonov-Bohm effect \cite{Aharonov:1959fk}. It might be considered as an indication of this quantum mechanical phenomenon at the classical level. We further demonstrate the equivalence of the \pu{null} displacement electromagnetic memory formula and the formula of sub-leading soft photon theorem. The new memory formula is just a Fourier transformation of the sub-leading soft factor.

The plan of this paper is quite simple. In next section, we will derive the formula of the new displacement electromagnetic memory after a short review of `kick' memory effect. Section \ref{s-soft} will present the connections between electromagnetic memory effects and soft photon theorems. Some comments on future directions will be given in the discussion section. There is also an appendix on the current of a collection of charged particles.

\section{Electromagnetic memory effect}
\label{memory}
We begin with a revisit on the general solutions of electromagnetic theory far from the source in Minkowski space-time. The retarded coordinates are applied which will only cover the future null infinity. However everything can be similarly repeated on the past null infinity in the advanced coordinates. The retarded spherical coordinates are connected to the Cartesian coordinates as follows:
\begin{equation}
\label{retard}
u=t-r,\;r=\sqrt{x^ix_i},\;x^1+ix^2=\frac{2rz}{1+z\bz},\;x^3=r\,\frac{1-z\bz}{1+z\bz}.
\end{equation}
The metric is
\begin{equation}
\label{metric}
\td s^2=-\td u^2-2\td u\, \td r+2r^2\gamma_{z\bz}\td z\td\bz,\;\; \gamma_{z\bz}=\frac{2}{(1+z\bz)^2}.
\end{equation}
The non-zero Christoffel symbols associated to \eqref{metric} are
\be
\Gamma^u_{z\bz}=r \ga\,,\;\;\;\;\Gamma^r_{z\bz}=-r\ga\,,\;\;\;\;\Gamma^z_{rz}=\frac{1}{r}\,,\;\;\;\;\Gamma^z_{zz}=\p\ln\ga\,.
\ee
We will work in the radial gauge $A_r^+=0$
\footnote{The plus sign denotes the fields at future null infinity.}. Once the asymptotic behavior of the gauge fields $A_\mu$ and the conserved current $j_\mu$ are specified in the following way \cite{Winicour:2014ska}\footnote{We have eliminated the Coulomb field of a moving charge which is not needed for our radiation memory investigation.}
\begin{equation}\label{condition}
\begin{split}
&A_u^+=\frac{A_u^{+0}(u,z,\bz)}{\gr{r}}+\cO(r^{\gr{-2}}),\\
&A_z^+=A_z^{+0}(u,z,\bz)+\frac{A_z^{+1}(u,z,\bz)}{r}+\cO(r^{-2}),\\
&j_u^+=\frac{j_u^{+0}(u,z,\bz)}{r^2}+\cO(r^{-3}),\\
&j_r^+=\cO(r^{-4}),\\
&j_z^+=\frac{j_z^{+0}(u,z,\bz)}{r^2}+\cO(r^{-3}),
\end{split}
\end{equation}
which defines an isolated electromagnetic system, the Maxwell's equations $\nabla_\nu F^{\mu\nu}=4\pi j^\mu$ yield \cite{Conde:2016csj}\footnote{In \cite{Conde:2016csj}, $j_r$ was set to be zero by the ambiguity of a conserved current. It is easy to find that \eqref{supeq+} and \eqref{puA1+} are still true when $j_r=\cO(r^{-4})$.}
\begin{align}
\label{supeq+}
  \p_u A^{+0}_u & =\gai \p_u(\p_z A^{+0}_{\bz} + \p_{\bz} A^{+0}_z) - 4\pi j^{+0}_u \,,\\
\label{puA1+}
  2\p_u A_z^{+1}&=\p_z A^{+0}_u  + \p_z [\gai(\p_z A^{+0}_{\bz} - \p_{\bz} A^{+0}_z)] - 4\pi j^{+0}_z \, .
\end{align}
As we will see later, \eqref{supeq+} and \eqref{puA1+} will completely determine the `kick' memory effect and the new (displacement) memory effect.

\subsection{`Kick' memory effect}
The `kick' memory effect was discovered by Bieri and Garfinkle in \cite{Bieri:2013hqa} (see also \cite{Winicour:2014ska}). Here we review the E mode memory effect for which we can set $A_{z(\bz)}^{+0}=\p_{z(\bz)} \alpha(u,z,\bz)$. Inserting this into \eqref{supeq+} leads to
\be\label{memory}
D_AD^A \delta \alpha=\delta A_u^{+0} + 4\pi \int^\infty_{-\infty}\td u\, j^{+0}_u \,,
\ee
where we have used the notation
\be
\delta F(z,\bz)=F(u,z,\bz)|_{u=\infty}-F(u,z,\bz)|_{u=-\infty},\nn
\ee
and $D_A$ is the spherical covariant derivative. The angular part of the electric fields are $E_{z(\bz)}=F_{z(\bz)u}$. They are related to the memory formula as
\be\label{kick}
\int^\infty_{-\infty}\td u E_{z(\bz)}^{\gr{0}}=- \p_{z(\bz)}\delta \alpha,
\ee
\gr{where $E_{z(\bz)}^0$ is the leading order of $E_{z(\bz)}$ in $\frac1r$.} From the standard analysis of the motion of a charged particle in the presence of electric fields, \eqref{kick} will leave a residual velocity to the charged particle (a `kick') \cite{Bieri:2013hqa}. Following the terminology of Bieri and Garfinkle, the first term on the right hand side of \eqref{memory} is called ordinary `kick' and the second one induces a null `kick'.

\subsection{Displacement memory effect}
For the B mode part where we can set $A_z^{+0}=i\p_z \beta(u,z,\bz)$ and $A_{\bz}^{+0}=-i\p_{\bz} \beta(u,z,\bz)$, it was proven that $\beta=0$ at $u=\pm\infty$ in the case of physically realistic source \cite{Winicour:2014ska}. Obviously $\Delta \beta=0$, which means no B mode `kick' memory though it is mathematically possible. Alternatively we will propose a new type of memory which is defined as
\be\label{new}
\delta \Gamma=\int^\infty_{-\infty}\td u\,\beta.
\ee
In terms of the electric fields, one will get
\be\label{dis}\begin{split}
&\int^\infty_{-\infty}\td u \int^u_{-\infty}\td u'\, E_{z}^{\gr{0}}=-i \p_{z}\delta \Gamma,\\
&\int^\infty_{-\infty}\td u \int^u_{-\infty}\td u'\, E_{\bz}^{\gr{0}}= i\p_{\bz}\delta \Gamma.
\end{split}\ee
Considering a very slowly moving charged particle, the electric field dominates the motion of the charged particle. Hence the new memory effect \eqref{dis} with one more integration over $u$ than the `kick' memory effect \eqref{kick} will be a displacement of the charged particle. By applying \eqref{puA1+}, the new memory formula can be arranged into a nicer way:
\be\label{newmemory}
i D_AD^AD_BD^B\delta\Gamma=\gr{2}\gai\delta(\p_z A_{\bz}^{+1}-\p_{\bz} A_z^{+1}) +\gr{4}\pi \gai \int^\infty_{-\infty}\td u\,(\p_z j_{\bz}^{+0}-\p_{\bz} j_z^{+0}).
\ee
According to the terminology of the `kick' memory, we will refer to the first term on the right hand side of \eqref{newmemory} as ordinary displacement and the second one as null displacement.

This new memory formula has a close relation to the formula of the Aharonov-Bohm effect \cite{Aharonov:1959fk}. To see that concretely, let us consider a circle $\cC$ at the future null infinity and define
\be
P=\frac{e}{\tau}\oint_{\cC}\,\left(A_z^{+0}\td z + A_{\bz}^{+0}\td \bz\right),
\ee
where $\tau$ and $e$ are respectively the period and the electric charge of the circular motion detector. So $P$ measures the rate of the phase change of a detector. Suppose $\tau$ is very small, the total phase shift will be essentially determined by the new memory effect as follow
\be
\int^\infty_{-\infty}\td u\, P=i\,\gr{\frac{e}{\tau}}\oint_{\cC}\,\left(\p_z\delta \Gamma\td z - \p_{\bz}\delta \Gamma\td \bz\right).
\ee

\subsection{Antipodal identification}
We can perform a similar analysis to get the memory formula near the past null infinity. There, the advanced coordinates
\begin{equation}
\label{advanced}
v=t+r,\;r=\sqrt{x^ix_i},\;x^1+ix^2=\frac{2rz}{1+z\bz},\;x^3=r\,\frac{1-z\bz}{1+z\bz},
\end{equation}
are needed which leads the metric to the following way:
\begin{equation}
\td s^2=-\td v^2+2\td v\, \td r+2r^2\gamma_{z\bz}\td z\td\bz.
\end{equation}
The analogues of \eqref{supeq+} and \eqref{puA1+} near past null infinity are
\begin{align}
\label{supeq-}  \p_v A^{-0}_v & =-\gai \p_v(\p_z A^{-0}_{\bz} + \p_{\bz} A^{-0}_z) + 4\pi j^{-0}_v \,,\\
\label{puA1-}  2\p_v A_z^{-1}&=\p_z A^{-0}_v  - \p_z [\gai(\p_z A^{-0}_{\bz} - \p_{\bz} A^{-0}_z)] + 4\pi j^{-0}_z \, .
\end{align}

The scattering problem involves both ingoing data at past null infinity $\skym$ and outgoing data at future null infinity $\skyp$. Hence a canonical identification between the far past of $\skyp$ and the far future of $\skym$ will be crucial in order to compare the memory formulas to the formulas of soft photon theorems in scattering process. Such identification was given for electromagnetic theory in \cite{He:2014cra,Lysov:2014csa}. According to their argument, it is natural to require
\be
A_{z(\bz)}^+(u\rightarrow-\infty)=A_{z(\bz)}^-(v\rightarrow\infty).
\ee
With this identification and the help of \eqref{supeq+}-\eqref{puA1+} and \eqref{supeq-}-\eqref{puA1-}, one can derive the total null memory formulas as\footnote{We have taken the convention of all legs outgoing.}\footnote{\pu{Since we only consider null memory, $A_u^{+0}|_{u\rightarrow+\infty}=A_v^{-0}|_{v\rightarrow-\infty}=0$ and $A_z^{+1}|_{u\rightarrow+\infty}=A_z^{-1}|_{v\rightarrow-\infty}=0$ should be noticed.}}
\be\label{l-memory}
4\pi\,\ga(\int^\infty_{-\infty}\td u\, j^{+0}_u + \int^\infty_{-\infty}\td v\, j^{-0}_v)=\p_z\Delta A_{\bz}+\p_{\bz}\Delta A_z,
\ee
and\footnote{\pu{One should notice that only imaginary part was involved in the definition of the new memory formula \eqref{new} and \eqref{newmemory}.}}
\be\label{sb-memory}
2\pi\, \text{Im}(\int^\infty_{-\infty}\td u\,\p_{\bz} j^{+0}_z + \int^\infty_{-\infty}\td v\,\p_{\bz} j^{-0}_z )=
\p_z\p_{\bz}\, \text{Im}(\gai\p_z\Delta {\mathbf\cA}_{\bz}),
\ee
where
\be\begin{split}
&\Delta A_{z(\bz)}=A_{z(\bz)}^{+0}(u\rightarrow+\infty)-A_{z(\bz)}^{-0}(v\rightarrow-\infty),\nn\\
&\Delta {\mathbf\cA}_{z(\bz)}=\int^\infty_{-\infty}\td u\,A_{z(\bz)}^{+0}+\int^\infty_{-\infty}\td v\,A_{z(\bz)}^{-0}.
\end{split}\ee
We will show in the next section that the total null memory formulas \eqref{l-memory} and \eqref{sb-memory} are nothing but the Fourier transformation of the formulas of leading and sub-leading soft photon theorems when the current of a collection of charged particles \eqref{u-current}-\eqref{z-currentv} is inserted.

\section{Equivalence to soft photon theorem}
\label{s-soft}
The fascinating connection between memory formula and soft theorem formula was first discovered by Strominger and Zhiboedov \cite{Strominger:2014pwa}. They found that the displacement gravitational memory formula is equivalent to Weinberg's formula for soft graviton emission \cite{Weinberg:1965nx}. Inspired by the new universal properties of soft graviton emission at the sub-leading order in the low-energy expansion \cite{Cachazo:2014fwa} (see also \cite{Gross:1968in,Jackiw:1968zza,White:2011yy} for earlier, more restricted, versions), a new gravitational memory was discovered in \cite{Pasterski:2015tva}. The new memory affects orbiting objects and it is called the spin memory effect. The spin memory formula was shown to be equivalent to the formula of sub-leading soft graviton theorem. In this section, we will illustrate the `kick' and displacement memory formulas in electromagnetic theory are respectively equivalent to the formula of leading and sub-leading soft photon theorems respectively.

Soft photon theorem \cite{Low:1954kd,GellMann:1954kc,Low:1958sn} (see also \cite{Weinberg:1965nx}) states that: In the process of $n\rightarrow m+1$ scattering where the $+1$ is a photon with very low energy (becoming soft), the tree-level scattering amplitudes have universal properties at leading and sub-leading order in a low-energy expansion:
\be
\label{soft}
M_{n+m+1}\big(p_1,\ldots,p_n,\left\{q;\epsilon^{\pm}\right\}\big)=
\left(\pu{\frac1\omega}S^{(0)\pm}+S^{(1)\pm}\right)M_{n+m}(p_1,\ldots,p_n)+\cO\left(\omega^1\right),
\ee
where
\begin{equation}
\label{S0}
  S^{(0)\pm}=\pu{\omega}\sum_{k=1}^{n+m} e_k\,\frac{p_k\cdot \epsilon^{\pm}}{p_k\cdot q}
\end{equation}
and
\begin{equation}
\label{S1}
  S^{(1)\pm}=\sum_{k=1}^{n+m} e_k\,\frac{q_{\mu}\epsilon_{\nu}^{\pm}\,J_k^{\mu\nu}}{p_k\cdot q} \ .
\end{equation}
$J^{\mu\nu}$ is angular momentum operator, $\omega=q^0$ and $\epsilon^{\pm}$ are respectively the energy and polarization vector of the soft photon\footnote{Here the plus and minus sign denote the helicity of the soft photon. One should not confuse with the ones denoting fields at different null infinities in the previous section.}.

\pu{We would like to understand a classical result as a limiting case of quantum result. A well known example is that the quantum calculation recovers the classical result of the radiated energy in the soft bremsstrahlung process \cite{Peskin:1995ev}. With such concept in mind,}
the key observation to see the equivalence between soft photon theorems and electromagnetic memory effects \cite{Strominger:2014pwa,Pasterski:2015tva,Pasterski:2015zua} is to consider the soft factors as the expectation value of asymptotic fields fluctuation:
\be
S^{(0)\pm}=\lim_{\omega\rightarrow0} \omega\frac{M_{n+m+1}^{\pm}}{M_{n+m}},
\ee
\be
S^{(1)\pm}=\lim_{\omega\rightarrow0}\p_\omega\left(\omega \frac{M_{n+m+1}^{\pm}}{M_{n+m}}\right).
\ee
Then \pu{the expectation value of asymptotic fields fluctuation with soft photon emission is simply related to the change of classical fields} under large-$r$ stationary phase approximation \cite{Strominger:2014pwa,Pasterski:2015tva,Pasterski:2015zua,He:2014laa,Kapec:2014opa} as
\pu{\be\Delta A_\mu \epsilon^{\mu\pm}=S^{(0)\pm},\quad\Delta {\mathbf\cA}_\mu \epsilon^{\mu\pm}=S^{(1)\pm}.\ee
Hence\footnote{\pu{Compared to \cite{Pasterski:2015zua}, the pre-factor $-\frac{e}{4\pi}$ is missing as we took a different convention in defining the current of charged particles \eqref{cu}.}}}
\be\label{leading}
\Delta A_z=\epsilon^{*+}_z S^{(0)+} + \epsilon^{\ast-}_z S^{(0)-},
\ee
\be\label{subleading}
\Delta {\mathbf\cA}_{z}=\epsilon^{*+}_z S^{(1)+} +\epsilon^{*-}_z S^{(1)-},
\ee
\pu{where $\epsilon^{*}$ is the complex conjugate of the polarization vector while the orthogonality condition $\epsilon_\mu^+\epsilon^{\mu+}=\epsilon_\mu^-\epsilon^{\mu-}=0$ and the normalization condition $\epsilon_\mu^+\epsilon^{*\mu+}=\epsilon_\mu^-\epsilon^{*\mu-}=1$ of the polarization vector have been used.}

\subsection{Leading order}

The null momenta by their energy and direction on the sphere \pu{in the Cartesian frame} can be parametrized as \pu{\cite{He:2014laa,He:2014cra}}
\begin{align}
\label{p}  p_k^{\,\pu{\mu}}&=E_k\left(1,\,\frac{w_k+\bar{w}_k}{1+w_k\bar{w}_k},\,i\frac{\bar{w}_k-w_k}{1+w_k\bar{w}_k},\,\frac{1-w_k\bar{w}_k}{1+w_k\bar{w}_k}\right),\\
\label{q}   q^{\pu{\mu}}&=\omega_q\left(1,\,\frac{w+\bar{w}}{1+w\bar{w}},\,i\frac{\bar{w}-w}{1+w\bar{w}},\,\frac{1-w\bar{w}}{1+w\bar{w}}\right).
\end{align}
Similarly one can parametrize polarization vectors as
\begin{equation}\begin{split}
\epsilon^{\pu{\mu}\,-}(q)=\frac{1}{\sqrt{2}}\left(\bar{w},1,-i,-\bar{w}\right),\\
\epsilon^{\pu{\mu}\,+}(q)=\frac{1}{\sqrt{2}}\left(w,1,i,-w\right).
\end{split}\end{equation}
Project the polarization vectors onto the sphere
\begin{equation}\begin{split}
&\epsilon_w^+(q)=\frac{\sqrt{2}}{1+w\bw}\,,\qquad \epsilon_{\bw}^+(q)=0\,,\\
&\epsilon_{w}^-(q)=0\,,\qquad\epsilon_{\bw}^-(q)=\frac{\sqrt{2}}{1+w\bw} \,.\end{split}
\end{equation}
by using $\epsilon_{w(\bw)}^\alpha=\frac1r \p_{w(\bw)} x^\mu\epsilon_{\mu}^\alpha$. With all those ingredients, we are ready to concrete the equivalence.

In the asymptotic position space, the leading soft factor reads:
\begin{align}
S^{(0)+}=\sum_{k=1}^{m+n} \frac{e_k}{\sqrt2}\frac{1+w\bw}{\bw-\bw_k},\quad
S^{(0)-}=\sum_{k=1}^{m+n} \frac{e_k}{\sqrt2}\frac{1+w\bw}{w-w_k}.
\end{align}
Inserting those quantities into \eqref{leading} (recall the complex conjugate), one gets
\be\label{1}
\Delta A_w=\sum_{k=1}^{m+n} e_k\frac{1}{w-w_k},\quad
\Delta A_{\bw}=\sum_{k=1}^{m+n} e_k\frac{1}{\bw-\bw_k}.
\ee
Acting with $\p_{\bw(w)}$ on $\Delta A_{w(\bw)}$ and considering the real part, we get\footnote{The relation $\p_z \frac{1}{\bz-\bz_k}=2\pi \delta^2(z-z_k)$ should be understood.}
\be
\p_{\bw}\Delta A_{w} + \p_w \Delta A_{\bw}=4\pi\sum_{k=1}^{m+n} e_k\,\delta^2(w-w_k),
\ee
which recovers precisely the result of inserting \eqref{u-current} and \eqref{u-currentv} into the memory formula \eqref{l-memory}.

\subsection{Sub-leading order}
In the asymptotic position space, the sub-leading soft factor is:
\be
S^{(1)+}=\sum_{k=1}^{m+n} \frac{e_k}{\sqrt2}\frac{b_k^3(1-w\bw_k)-i b_k^2(w-\bw_k)+b_k^1(w+\bw_k)}{\bw-\bw_k},
\ee
\be
S^{(1)-}=\sum_{k=1}^{m+n} \frac{e_k}{\sqrt2}\frac{b_k^3(1-\bw w_k)\pu{+}i b_k^2(\bw-w_k)+b_k^1(\bw+w_k)}{w-w_k},
\ee
where one needs to identify the angular momentum operator to the angular momentum of a collection of particles \eqref{angulr}. From \eqref{subleading}, one gets
\be
\Delta{\mathbf\cA}_{w}=\sum_{k=1}^{m+n} e_k \\ \frac{b_k^3(1-\bw w_k)-i b_k^2(\bw-w_k)+b_k^1(\bw+w_k)}{(1+w\bw)(w-w_k)},
\ee
\be
\Delta{\mathbf\cA}_{\bw}=\sum_{k=1}^{m+n} e_k \\ \frac{b_k^3(1-w\bw_k)\pu{+}i b_k^2(w-\bw_k)+b_k^1(w+\bw_k)}{(1+w\bw)(\bw-\bw_k)}.\label{2}
\ee
Acting with $\p_{w(\bw)}\p_{\bw(w)}\left(\gawi \p_{w(\bw)}\right)$ on $\Delta \mathbf\cA_{\bw(w)}$ and taking the imaginary part, one obtains\footnote{\pu{The relation $w \delta^2(w-w_k)=w_k \delta^2(w-w_k)$ will be used.}}
\begin{multline}
\text{Im}\,\p_w\p_{\bw}(\gawi\p_w \Delta \mathbf\cA_{\bw})=\pi\,\text{Im}\sum_{k=1}^{m+n} e_k\\ \left[b_k^1(1-\bw_k^2)-i b_k^2(1+\bw_k^2)-2b_k^3\bw_k\right]\p_{\bw}\delta^2(w-w_k),
\end{multline}
which gives exactly the same expression as injecting \eqref{z-current} and \eqref{z-currentv} into \eqref{sb-memory}.

\section{Discussions}
\pu{In this work, we have proposed a new electromagnetic memory. The new memory effect is a displacement of a charged particle. Hence the possible detection of this effect should be much simpler than the detection of the `kick' electromagnetic memory suggested in \cite{Susskind:2015hpa}. There are two types of displacement electromagnetic memory, namely the ordinary and null displacement which are generated by massive and massless charged matters respectively. We have shown precisely the equivalence between the null electromagnetic memory formulas (`kick' memory and displacement memory) and the soft photon theorems (leading and sub-leading soft photon theorems). The formers are nothing but a Fourier transformation of the laters.}

The displacement memory effect is a magnetic type, or B mode, radiation memory. Remarkably a careful analysis shows that the new gravitational memory proposed in \cite{Pasterski:2015tva} is also magnetic type though the displacement gravitational memory can not be derived in purely asymptotic argument \cite{Winicour:2014ska}. An interesting result of \cite{Cachazo:2014fwa} is that the soft graviton theorems go through the third order in the low-energy expansion. It is very curious to ask if there should be a third gravitational memory. A recent investigation on linearized gravity theory \cite{Conde:2016rom} gives positive hint on the possibility of a third gravitational memory effect. It would be very meaningful to derive the concrete formulas elsewhere \cite{2come}.

\section*{Acknowledgments}

The authors thank Glenn Barnich and Eduardo Conde for useful discussions, and additionally Eduardo Conde again for valuable comments on the draft. This work is supported in part by the National Natural Science Foundation of China (Grant Nos. 11575202, 11575286 and 11475179).

\appendix

\section{Massless charged particle current}

We will present the current of a collection of charged particles which is needed for computing the memories. The trajectory of massless point particle is \cite{Pasterski:2015tva}
\be
x^\mu(\tau)=\frac{p^\mu}{E}\tau + b^\mu,
\ee
where $b^\mu=(0,b^i)=x^{\gr{\mu}}(0)$ is the particle trajectory relative to the space-time origin. The orbital angular momentum of this particle is
\be\label{angulr}
L^{\mu\nu}=b^\mu p^{\pu{\nu}} - b^\nu p^\mu.
\ee
For simplicity, we will consider particle without intrinsic spin. However the way to include intrinsic spin is straightforward. The asymptotic behavior of the trajectory at large $\tau$ in retarded coordinates \eqref{retard} is as follows
\be
\begin{split}
&r(\tau)\pu{=\sqrt{x^ix_i}}=\tau+\frac{p^\mu}{E^2}L_{u\mu}+\cO (\tau^{-1}),\\
&u(\tau)\pu{=t-\sqrt{x^ix_i}}=-\frac{p^\mu}{E^2}L_{u\mu}+\cO (\tau^{-1}),\\
&z(\tau)\pu{=\frac{\sqrt{x^ix_i}-x^3}{x^1-i x^2}}=\frac{p^1+i p^2}{E+p^3}+\frac{L_{u\bz}(z,\bz)}{E\tau\ga}+\cO (\tau^{-2}),
\end{split}
\ee
where
\be\begin{split}
L_{u\bz}(z,\bz)&\equiv\lim_{r\rightarrow\infty}\frac{1}{r}\frac{\p x^\mu}{\p u}\frac{\p x^\nu}{\p \bz}L_{\mu\nu}\\
&=\frac{b^1(1-z^2)+i b^2(1+z^2)-2b^3z}{(1+z \bz)^2}E.\end{split}
\ee
The current of the point particle is defined as
\be\label{cu}
j^\mu(y^\rho)=\pu{-}\,e \int \td \tau \,\dot{x}^\mu\,\frac{\delta^4(y^\rho-x^\rho(\tau))}{\sg},
\ee
where $e$ is the charge of the particle. For a collection of free point particles, one just needs to sum over. The asymptotic behavior of current is
\begin{align}
&\lim_{r\rightarrow\infty}r^2 j_u=\sum_k e_k\,\delta(u-u_k)\frac{\delta^2(z-z_k)}{\ga},\label{u-current}\\
&\lim_{r\rightarrow\infty}r^2 j_z=\sum_k \frac{e_k}{E_k}\,\delta(u-u_k)L_{uz_k}(z_k,\bz_k)\frac{\delta^2(z-z_k)}{\ga},\label{z-current}
\end{align}
and $j_r=\cO(r^{-4})$.

Similarly, one obtains \pu{(recall the convention of all legs outgoing)}
\begin{align}
&\lim_{r\rightarrow\infty}r^2 j_v=\sum_k e_k\,\delta(v-v_k)\frac{\delta^2(z-z_k)}{\ga},\label{u-currentv}\\
&\lim_{r\rightarrow\infty}r^2 j_z=\sum_k \frac{e_k}{E_k}\,\delta(v-v_k)L_{uz_k}(z_k,\bz_k)\frac{\delta^2(z-z_k)}{\ga},\label{z-currentv}
\end{align}
and $j_r=\cO(r^{-4})$ in advanced coordinates.

\bibliography{ref}

\bibliographystyle{utphys}

\end{document}